\documentclass[a4paper,11pt]{article}
\usepackage{pos}
\usepackage{float}
\usepackage{lineno}

\title{Unraveling the nature of GRBs progenitors through neutrinos}
 \ShortTitle{GRBs progenitors though neutrinos}

\author*[a]{Gibran Morales}
\author[a]{N. Fraija}

\affiliation[a]{Instituto de Astronom\'ia, Universidad Nacional Aut\'onoma de M\'exico, Ciudad de Mexico, Mexico}

\emailAdd{gmorales@astro.unam.mx }
\emailAdd{nifraija@astro.unam.mx}

\abstract{GRBs are the most energetic electromagnetic events in the Universe. Those whose typical duration is longer than a few seconds are known as long GRBs and shorter than a few seconds are short GRBs. It is widely accepted that these events are associated with the collapse of a very massive star and the neutron star (NS) binary merger, respectively. A fast-spinning, strongly magnetized NS could be expected before a black hole (BH) in both scenarios. We allude to the thermal neutrinos’ particular properties propagating inside the fireball for differentiating both scenarios in this work. We first derive the neutrino effective potential associated with each medium in a strong and weak magnetic field. We calculate the three-flavor oscillation probabilities, and finally, we get the expected neutrino rate in both scenarios. Given these observables’ evolution, we can determine whether the progenitor could be associated with a strongly magnetized NS or a BH.}

\FullConference{37$^{\rm{th}}$ International Cosmic Ray Conference (ICRC 2021)\\
		July 12th -- 23rd, 2021\\
		Online -- Berlin, Germany}

\begin{document}
\maketitle
\section{Introduction}
	\paragraph*{}
Inside a GRB, isotropic luminosities of up to $10^{54}$ erg are reached in a tiny fraction of time.
During several years of observation, these GRBs were found to exhibit a bimodal distribution according to the duration of their early emission, with a marked separation around two seconds. This served to classify them into short (SGRBs) and long gamma-ray bursts (LGRBs) \citep {berger2014short}. This bimodality suggests the existence of two distinct classes of progenitors. In the case of SGRBs, short timescales (on the order of milliseconds) suggest a model of a progenitor based on the merger of two compact objects, such as neutron star-neutron star or black-hole-neutron star \citep{ lee2007progenitors}.  On the one hand, the progenitors of LGRBs are associated with the collapse of a massive star leading an Ic-type supernova (CCSNe), based on the exclusive location in galaxies with active star formation and by a strong correlation with UV-bright regions in their host galaxies \citep {woosley1993gamma, paczynski1998gamma, macfadyen1999collapsars}. There are two scenarios associated with the collapse of massive stars. In the first scenario, the collapse of massive stars typically leads to a BH plus a long-lived debris torus system \citep[e.g., see][]{1993ApJ...405..273W, 1977MNRAS.179..433B} and  in the second scenario, a millisecond magnetar is created with sufficient rotational energy to prevent the gravitational collapse \cite[e.g., see][]{2011MNRAS.413.2031M}. In both cases, a spinning disk of debris may be left in the vicinity of the compact item. Because the temperature is higher than the rate of positron/electron pair formation, nuclei are photo-disintegrated, and the plasma is mostly composed of free pairs, gamma-ray photons, and baryons. The base of the jet is created by the so-called fireball plasma linked to the progenitor.

However, the study of GRBs is far from done, with many unanswered questions. In that sense, neutrinos represent a worthwhile opportunity to characterize the remaining (and sometimes hidden) GRB sources, as many of them are produced in the newly born post-merger disc. This is especially relevant because recent work has discussed the possibility of detecting thermal neutrinos from these sources on future Megaton detectors. Although neutrinos belong to the least sensitive channel compared to electromagnetic and gravitational ones, they may provide additional information to confine the progenitor's position if the medium to photons is opaque. In this work, we present the neutrino effective potential in a magnetized fireball and calculate the three-flavor oscillation probabilities in a strong and weak magnetic field scenario. We are solely interested in those created  by thermal processes in the MeV range. The characteristics of these neutrinos change as they propagate through a non-vacuum medium due to the Mikheyev-Smirnov-Wolfenstein effect, therefore these extra consequences must be considered in the study of neutrino behavior. Finally, we get the expected neutrino rate in both scenarios of CCSNe.


\section{Neutrino Oscillation and propagation}\label{}	
	\paragraph*{}
	
	Neutrinos are weakly interacting particles that come in three flavors, each of which corresponds to a leptonic family. They oscillate with each other, and their transitions can be discussed in vacuum using analytical formulas. L. Wolfenstein demonstrated in 1978 that neutrinos propagating in a medium other than vacuum are affected by an effective potential equal to the medium's refractive index. Later, S. Mikheyev and A. Smirnov  \citep{mikheyev1986yad} demonstrated that neutrino oscillation parameters do change when neutrinos propagate through a material medium. The MSW mechanism is the name given to this theory. This additional potential increases the effective mass of the neutrinos, as well as their mass and flavor eigenstates; the conditions of this medium must be incorporated into the calculation of each allowed probability, which quickly becomes complicated.
	
\subsection{Neutrino production mechanisms}	
Due to the high temperatures reached  within the collapse, many neutrino emission mechanisms occur inside the plasma fireball, the main processes being:

\citep{lat76}:
\begin{itemize}
\item pairs annihilation ($e^++e^-\to\nu_x+\bar{\nu}_x$),
\item plasmon decay $(\gamma\to\nu_x+\bar{\nu}_x)$,
\item photo-neutrino emission $(\gamma+e^-\to e^-+\nu_x+\bar{\nu}_x)$,
\item positron capture $(n+e^+\to p+\bar{\nu}_e)$,
\item electron capture $(p+e^-\to n+\nu_e)$,
\end{itemize}

Since the final interaction is the only one that generates neutrinos with a specified electron flavor, therefore we will suppose that the starting neutrino rate in this study is $(4\ \nu_e:3\ \nu_\mu:3\ \nu_\tau)$
	
\subsection{Neutrino potentials within a magnetized fireball}
In this work we use the neutrino potential for a magnetized fireball in two regimes derived by \citep{fra14}. The first one with a strong magnetic field, and the second case with a weak magnetic field.
\subsection*{Strong $\vec{B}$ limit}

\begin{eqnarray}\label{eq:veffs}
V_{\rm eff,s}=\frac{\sqrt2\,G_F\,m_e^3 B}{\pi^2\,B_c}\biggr[\sum^{\infty}_{l=0}(-1)^l\sinh\alpha_l   \left[F_s-G_s\cos\varphi \right]\nonumber\\
-4\frac{m^2_e}{m^2_W}\,\frac{E_\nu}{m_e}\sum^\infty_{l=0}(-1)^l\cosh\alpha_l  \left[J_s-H_s\cos\varphi \right]  \biggr]\,,\cr
 \end{eqnarray}
\subsection*{Weak $\vec{B}$ limit}
\begin{eqnarray}\label{eq:veffw}
V_{\rm eff,w}=\frac{\sqrt2\,G_F\,m_e^3 B}{\pi^2\,B_c}\biggr[\sum^{\infty}_{l=0}(-1)^l\sinh\alpha_l  \left[F_w-G_w\cos\varphi \right]\nonumber\\
-4\frac{m^2_e}{m^2_W}\,\frac{E_\nu}{m_e}\sum^\infty_{l=0}(-1)^l\cosh\alpha_l \left[J_w-H_w\cos\varphi \right]\,.\cr 
\end{eqnarray}
where $m_e$ is the electron mass, $\alpha_l=(l+1)\mu/T$  with $\mu$ and $T$ the chemical potential and temperature, respectively, $B_c=m_e^2/e=4.141\times 10^{13}\, {\rm\ G}$ is the critical magnetic field,  $E_\nu$ is the neutrino energy and the functions F$_s$, G$_s$, J$_s$, H$_s$ are  described in more detail in (\citep{fra14}) and are dependent on the parameters of the medium considered, such as, magnetic field, chemical potential, among others.
\section{Results and Conclusion}\label{sec:ConADis}
	\paragraph*{}
The goal of this manuscript is to investigate the conditions that the oscillation properties will have when passing through two different media, the first of which is provided by a magnetized fireball as a result of the remnant magnetar's contribution. In this case, typical $B=10^{15} $ G values are obtained. In the second case, we will investigate a fireball with only a minor magnetic contribution ($B=10^{12} $ G) from the neutron star precursor of the LGRBs.

Using Equations (\ref{eq:veffs}) and (\ref{eq:veffw}), we show how this potential behaves as a function of temperature in both regimes. We discover that this potential is an increasing function with a range of $3\times10^{-9}$ to $7\times10^{-7}$ eV and that it is indeed dependent on the magnetic field strength under consideration. We can begin calculating the probabilities of oscillation once we have a reliable expression describing the medium through which these neutrinos travel. As previously stated, neutrinos come in three flavors, each with six independent oscillation probabilities. In Figure \ref{fig:P_E}, we only show two of them as a function of neutrino energy. We can see that they deviate from the expected theoretical value and that the magnetic field influences their values.

We can even go a step further and plot how these probabilities change as the fireball parameters change. As a result, in both cases, we see the oscillation adjusting different medium temperature parameters in Figure 2.	Because of ground-based detectors cannot quantify oscillation probabilities, we must also compute the predicted neutrino rate of these occurrences. For these, we use a well-known parameterization to derive the flavor rates as a function of energy, where neutrino fraction are described as  $\xi_n=F_n/\sum_n F_n$, which is what we would expect to see. When we compare them to the theoretical rates anticipated in a vacuum, we notice that the two circumstances differ. Finally, combining these findings would allow us to determine the initial magnetic field conditions that the progenitors experienced during neutrino production, for example. This would be used as a secondary detection channel to determine whether the central remnant is a magnetar or simply a neutron star.

\begin{figure*} 
\centering
\includegraphics[width=0.56\textwidth]{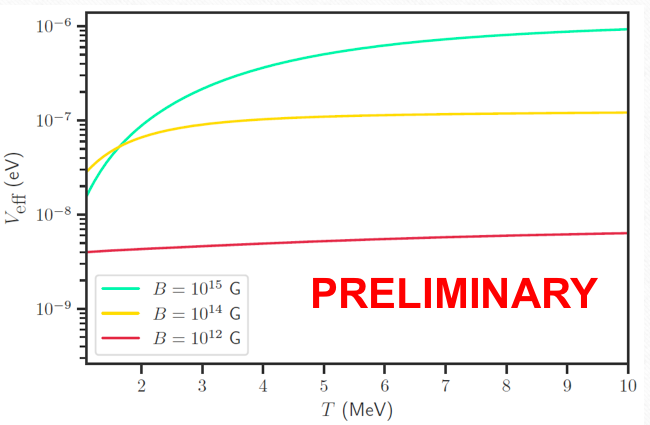}	
\caption{Representation of the neutrino potential derived by Equations (\ref{eq:veffs}) and (\ref{eq:veffw}) under multiple conditions inside a fireball. Particularly, in this plot we see the dependencies of this potential as a function of the temperature of the medium $T$. Here we have considered a magnetized fireball by the contribution of a neutron star with typical values of ($10^{12}$ G) and of a magnetar whose surface intensity is between ($10^{14}-10^{15}$) G.}
\label{fig:V_T}        	        
\end{figure*}

\begin{figure*}
    \centering
    \includegraphics[width=0.97\textwidth]{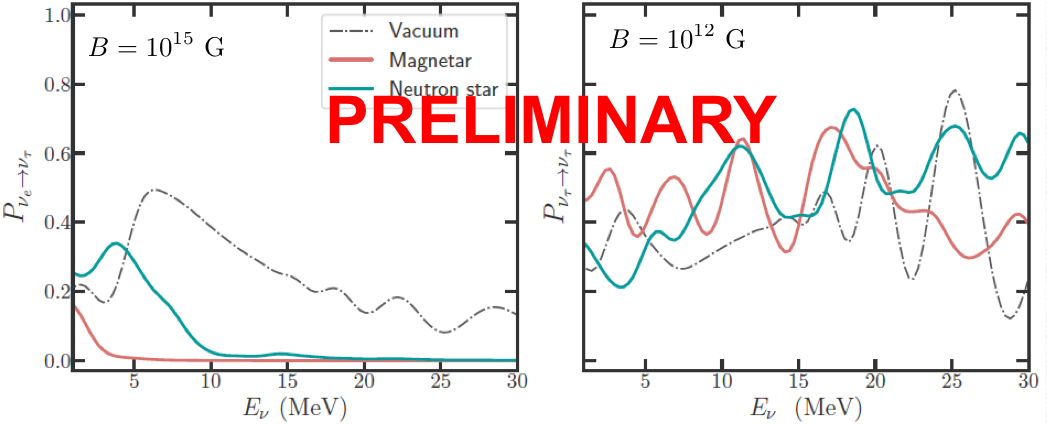}
    \caption{$P(\nu_e\to\nu_\tau)$ and $P(\nu_\tau\to\nu_\tau)$ oscillation probabilities for neutrinos travelling through a fireball with magnetic contributions from an isolated neutron star (blue line) and a magnetar projected onto the $P-E\nu$ plane (red line).  For comparison, we have also plotted the oscillations in the vacuum (dotted line). A Normal Ordering for neutrino mass hierarchy  scheme was considered for these plots.}
    \label{fig:P_E}
\end{figure*}

\begin{figure*}\centering
\includegraphics[width=0.97\textwidth]{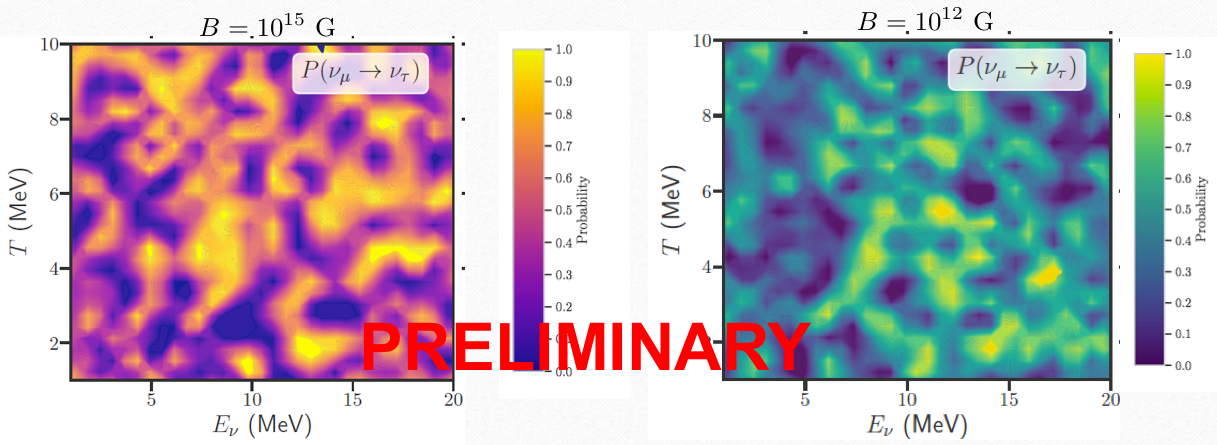}	

\caption{Contour plots of the expected oscillation probability for various medium-temperature and neutrino energy levels. Again, we investigate the magnetic contributions from a magnetar (left) and a neutron star (right). For this arrangement, we present the transition probability of $P(\nu_\mu\to\nu_\tau)$. It is worth noting that the electron neutrino's survival probability dominates in both scenarios but in differing proportions.}
\label{fig:T_E} 
\end{figure*}

\begin{figure*} 
\centering
\includegraphics[width=0.97\textwidth]{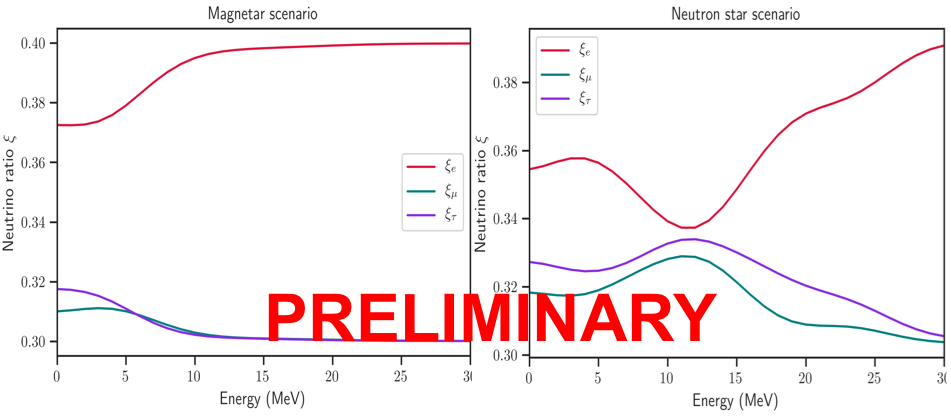}
\caption{Representation of the expected neutrino flavor fractions $\xi_i$  within a magnetar (left) and a neutron-star (right) scenario as a function of $E_\nu$. We find a greater contribution of electronic flavour due to an excess of $\nu _e$ produced by charged-current interactions ($e^\pm$capture on nucleons) within the medium.}
\label{fig:xi_E}        	        
\end{figure*}

\acknowledgments{
GM acknowledges the financial support through the CONACyT grant 825482. NF acknowledges the financial support from
UNAM-DGAPA-PAPIIT through IN106521.}	

\bibliographystyle{JHEP} 
\bibliography{Bib_osc}

\end{document}